\documentclass[12pt,superscriptaddress,showpacs,aps,preprint]{revtex4}

\usepackage{amsmath}
\usepackage{amssymb}
\usepackage{graphicx}
\usepackage{color}
\usepackage{subfigure}

\makeatletter

\usepackage{slashed}

\newcommand{\be}{\begin{equation}}
\newcommand{\ee}{\end{equation}}
\newcommand{\en}{\end{equation}}
\newcommand{\ba}{\begin{eqnarray}}
\newcommand{\ea}{\end{eqnarray}}
\newcommand{\bea}{\begin{eqnarray}}
\newcommand{\eea}{\end{eqnarray}}

\newcommand{\bq}{\begin{eqnarray}}
\newcommand{\eq}{\end{eqnarray}}

\makeatother

\begin{document}

\title {Modified Euler-Heisenberg effective action and Proper-Time Method in Lorentz-Violating Scalar QED}

\author{L. C. T. Brito}
\email{lcbrito@ufla.br}
\affiliation{Departamento de F\'{i}sica, Instituto de Ci\^{e}ncias Naturais,  Universidade Federal de Lavras, Caixa Postal 3037,
	37200-900, Lavras, Minas Gerais, Brasil}

\author{J. C. C. Felipe}
\email{jeanccfelipe@ufsj.edu.br}
\affiliation{Universidade Federal de S\~ao Jo\~ao del Rei, Departamento de Estat\'istica, F\'isica e Matem\'atica, Rod. MG-443, Km 7 - 36497-899 - Fazenda do Cadete - Ouro Branco, Minas Gerais, Brazil}

\author{A. C. Lehum}
\email{lehum@ufpa.br}
\affiliation{Faculdade de F\'{i}sica, Universidade Federal do Par\'{a}, 66075-110, Bel\'{e}m, Par\'a, Brazil}

\author{J. R. Nascimento}
\email{jroberto@fisica.ufpb.br}
\affiliation{Departamento de F\'{i}sica, Universidade Federal da Para\'{i}ba, Caixa Postal 5008,
	58051-970 Jo\~{a}o Pessoa, Para\'iba, Brazil}

\author{A. Yu. Petrov}
\email{petrov@fisica.ufpb.br}
\affiliation{Departamento de F\'{i}sica, Universidade Federal da Para\'{i}ba, Caixa Postal 5008,
	58051-970 Jo\~{a}o Pessoa, Para\'iba, Brazil}


%%%%%%%%%
\begin{abstract}
  Quantum photon effects in vacuum provide an interesting setting to test quantum electrodynamics, serving as a source for predictions about physics beyond the Standard Model. In this paper, we investigate these effects by calculating the one-loop Euler-Heisenberg-like effective action within a Lorentz-violating scalar quantum electrodynamics framework. In both CPT-even and CPT-odd scenarios, we obtain the exact result in all orders of the stress tensor $F_{\mu\nu}$ and evaluate explicitly the lower orders of this effective action. We identify the quantum effects coming from Lorentz violation in an explicitly gauge invariant way. Nonlinear Lorentz-violating contributions that may affect photon-photon scattering are explicitly evaluated.
\end{abstract}

\maketitle

%%%%%%%%%
%\section{Introduction}
%\label{I}
%%%%%%%%%

\section{Introduction}

Within studies of QED and its various extensions, an important role is played by calculations of the Euler-Heisenberg (EH) effective action involving all orders in the background stress tensor $F_{\mu\nu}$ and describing the one-loop photon-photon scattering. Originally, it was introduced in \cite{EH}. Further, many its generalizations have been proposed for various extensions of QED (see \cite{Dunne,Dunne:2012vv} for a review). 

One of the most efficient manners for its calculation is the proper-time method \cite{Schwinger} which has been generalized and used in various contexts including gravity (see e.g. \cite{BV}), superfield supersymmetry (see e.g. \cite{BKY,McA}) and many other contexts. Therefore, application of this method in the context of Lorentz-violating (LV) theories, especially, the Lorentz-violating Standard Model Extension (LV SME), formulated in \cite{ColKost1,ColKost2} is certainly a  interesting problem.

 First step in this direction has been done in \cite{ptYM}, where the non-Abelian Carroll-Field-Jackiw (CFJ) term and the gravitational Chern-Simons term have been calculated explicitly in the one-loop approximation from coupling of gauge theory and gravity respectively with the spinor matter in the presence of a LV background. Further, the EH effective action has been obtained for the first time with use of the proper-time method in \cite{FerrEH}, for a simplest situation where the impact of the LV terms consists in a simple modification of the background gauge field. Within this calculation, the perturbative calculations were based on using the version of the proper-time method discussed in \cite{McA}. Further, the EH effective action has been obtained in \cite{ourproperLV} in LV spinor QED for the cases of presence of LV terms described constant parameters of SME $a_{\mu},b_{\mu},c_{\mu\nu}$ (see \cite{ColKost1,ColKost2,KosPic}). It worth to mention also the explicit calculation of lower contributions to the EH effective action performed in \cite{Fur}.

 Although the proper time has been applied in an interesting variety of Lorentz-violating scenarios, there is another important sector of LV SME where the method has not been explored: the LV scalar QED.  In fact,  within it, we find the lower one-loop quantum corrections have been obtained in \cite{scalQED1,scalQED2,scalQED2a} for the CPT-even sector of the theory, and in \cite{scalQED3} for the CPT-odd one. In all these cases, the Feynman diagram method was employed. So, it is interesting to explore the proper-time method to study the EH effective action in this theory, where the use of diagrammatic methods may pose particular challenges for this calculation. This is the aim of this paper.

The structure of the paper looks as follows. In section 2, we evaluate the EH effective action in the CPT-even LV scalar QED, in section 3, we perform the analogous calculations in the CPT-odd LV scalar QED, and in section 4 we conclude with a brief discussion of our results.

\section{EH effective action in the CPT-even LV QED}

So, let us begin with considering the CPT-even version of LV QED:
\bea
{\cal L}=(D_{\mu}\phi)^{\dagger}(\eta^{\mu\nu}+c^{\mu\nu})D_{\nu}\phi-m^2\phi\phi^*-V(\phi\phi^*)-\frac{1}{4}F_{\mu\nu}F^{\mu\nu},
\eea
where $D_{\mu}=\partial_{\mu}-ieA_{\mu}$ id the gauge covariant derivative. Our aim will consist in obtaining the Heisenberg-Euler effective action, i.e. we integrate over the scalar field and obtain the one-loop effective action in all orders of the stress tensor $F_{\mu\nu}$, in the form of the following functional trace:
\bea
\Gamma^{(1)}[A]=i{\rm Tr}\ln (-D^2-c^{\mu\nu}D_{\mu}D_{\nu}-m^2).
\eea
This expression is a natural extension of that one discussed in \cite{McA}. We note that, since the scalar field is integrated out, its self-interaction does not generate any nontrivial impact.  The factor $i$ instead of $i/2$ is caused by the fact that we integrate over the complex scalar rather than the real one.

To evaluate this trace, we employ the proper time method. This is a powerful tool allowing to calculate loop corrections, being especially used for the one-loop approximation. The essence of this method looks like follows.

Let us consider some theory whose action is $S[\Phi]$ where $\Phi$ denotes the set of all dynamical fields in the theory. To calculate the one-loop effective action, we represent these fields as a sum of the background (classical) and quantum parts through the replacement $\Phi\to\Phi+\phi$, where now $\Phi$ being the background field, and $\phi$, the quantum one. Using the functional integral approach, we can show that the one-loop effective action looks like $\Gamma^{(1)}=\frac{i}{2}{\rm Tr}\ln S^{\prime\prime}[\Phi]$, where $S^{\prime\prime}[\Phi]$, a second functional derivative of the classical action, is an operator, e.g. in the scalar field theory with the Lagrangian ${\cal L}=-\frac{1}{2}\partial_m\phi\partial^m\phi-V(\phi)$ one has $S^{\prime\prime}[\Phi]=\Box+V^{\prime\prime}(\Phi)$, thus, $\Gamma^{(1)}=\frac{i}{2}{\rm Tr}\ln (\Box+V^{\prime\prime}(\Phi))$. To calculate this trace, we use the Schwinger proper time representation: $\ln A=\frac{1}{i}\int_0^{\infty}\frac{ds}{s}e^{isA}$, which in the present case looks like
$$
\frac{i}{2}{\rm Tr}\ln (\Box+V^{\prime\prime}(\Phi))=\frac{i}{2}{\rm Tr}\int_0^{\infty}\frac{ds}{s}e^{is(\Box+V^{\prime\prime}(\Phi)}.
$$
This expression can be easily generalized for more complicated cases and treated with use of different techniques. That one we use in the paper is described in \cite{McA}.

The advantage of this method, in comparison with the standard Feynman diagrams approach, consists in the fact that within the proper time approach all contributions from different Feynman diagrams are summed automatically, therefore, we can easily evaluate the result involving all orders in the background field. This approach is especially powerful when the contributions from the gauge sector are evaluated (cf. \cite{McA}), that is, just the case considered in our paper.

As the Lorentz symmetry breaking is small, this expression can be written down as a power series in $c_{\mu\nu}$, and we take into account only zero and first orders:
\bea
\Gamma^{(1)}[A]=i{\rm Tr}\ln (-D^2-m^2)+i{\rm Tr} \left[c^{\mu\nu}D_{\mu}D_{\nu}\frac{1}{D^2+m^2}\right].
\eea
The first term in this expression is the Lorentz-invariant contribution which has been evaluated explicitly in \cite{McA}. So, let us concentrate on the second term, describing the lower LV contribution:
\bea
\Gamma^{(1)}_{1,LV}[A]&=&i{\rm Tr} \left[c^{\mu\nu}D_{\mu}D_{\nu}\frac{1}{D^2+m^2}\right]=
i{\rm Tr}\int_0^{\infty} ds c^{\mu\nu}D_{\mu}D_{\nu} e^{-sD^2} e^{-sm^2}=\nonumber\\
&=&i\int d^4x\int_0^{\infty} ds e^{-sm^2} c^{\mu\nu}D_{\mu}D_{\nu} e^{-sD^2} \delta^4(x-x')|_{x=x'}.
\eea
Actually, we must calculate the trace $D_{\mu}D_{\nu} e^{-sD^2} \delta^4(x-x')|_{x=x'}$. It has been found in \cite{ourproperLV}:
\bea
\label{trace2d}
D_{\mu}D_{\nu} e^{-sD^2} \delta^4(x-x')|_{x=x'}=\left(\frac{-ieF}{e^{-2iesF}-1}\right)_{\nu\mu}\frac{1}{16\pi^2s^2}{\rm det}^{1/2}\left(\frac{iesF}{\sinh (iesF)}\right).
\eea
As a result, our lower LV contribution to the Euler-Heisenberg effective action is
\bea
\Gamma^{(1)}_{1,LV}[A]&=&i\int d^4x\int_0^{\infty} \frac{ds}{16\pi^2s^2} e^{-sm^2} c^{\mu\nu}\left(\frac{-ieF}{e^{-2iesF}-1}\right)_{\mu\nu}
{\rm det}^{1/2}\left(\frac{iesF}{\sinh (iesF)}\right).
\eea
This the exact analogue of the lower LV contribution in the spinor QED found in \cite{ourproperLV}.
Different orders in $F_{\mu\nu}$ can be read off from this expression through its expansion in power series. Extracting second- and fourth-order terms in the perturbation expansion, we arrive at:

\begin{equation}
-i\Gamma_{LV,1}^{(1)}[A]=e^{2}\Gamma_{a,1}^{(1)}+e^{4}\Gamma_{b,1}^{(1)},
\end{equation}
where 
\begin{equation}
\Gamma_{a,1}^{(1)}=-\frac{I_{1}(m^{2})}{32\pi^{2}}\int d^{4}x\left(\frac{1}{12}c_{\,\,\,\alpha}^{\alpha}F_{\mu\nu}F^{\mu\nu}-\frac{1}{3}c_{\,\,\,\nu}^{\mu}F_{\mu\rho}F^{\rho\nu}\right)
\end{equation}
is the quadratic contribution to the one-loop effective action, and
\begin{eqnarray}
\Gamma_{b,1}^{(1)}&=&\frac{I_{2}(m^{2})}{2304\pi^{2}} \int d^{4}x\left[c_{\,\,\,\alpha}^{\alpha}\left(\frac{1}{5}F_{\mu\sigma}F^{\sigma\rho}F_{\rho\nu}F^{\nu\mu}+\frac{1}{4}\left(F_{\mu\nu}F^{\mu\nu}\right)^{2}\right) \right. \nonumber \\
&+&\left. 2  \left(F_{\mu\nu}F^{\mu\nu}\right)\left(c^{\sigma\rho}F_{\sigma\lambda}F_{\,\,\rho}^{\lambda}\right)
-  \frac{192}{95}\,c^{\mu\nu}F_{\mu\sigma}F^{\sigma\rho}F_{\rho\lambda}F_{\,\,\,\nu}^{\lambda}\right],
\end{eqnarray}
is the quartic one. Here the $I_1$ and $I_2$ are the proper-time integrals given by:
\begin{equation}
I_{1}(m^{2})=\mu^{-2\epsilon}\int_{0}^{\infty}ds\frac{e^{-sm^{2}}}{s^{1-\epsilon}}=\frac{1}{\epsilon}+\ln\frac{m^2}{\mu^2}
\end{equation}
and
\begin{equation}
 I_{2}(m^{2})=\int_{0}^{\infty} ds e^{-sm^{2}}s=\frac{1}{m^4}.
\end{equation}
Here we introduced the $\epsilon$ regularization parameter in the only divergent integral $I_1(m^2)$. Taking all together, we write the quadratic contribution to the effective action as
\bea
\Gamma_{a,1}^{(1)}=-\frac{1}{32\pi^{2}}(\frac{1}{\epsilon}+\ln\frac{m^2}{\mu^2})\int d^{4}x\left(\frac{1}{12}c_{\,\,\,\alpha}^{\alpha}F_{\mu\nu}F^{\mu\nu}-\frac{1}{3}c_{\,\,\,\nu}^{\mu}F_{\mu\rho}F^{\rho\nu}\right),
\eea
with the divergence can be removed by an adding the corresponding counterterm.  We note that, if we introduce $c^{\mu}_{\nu}=\tilde{c}^{\mu}_{\nu}+\frac{1}{4}c_0\delta^{\mu}_{\phantom{\mu}\nu}$, with $\tilde{c}^{\mu}_{\phantom{\mu}\nu}$ is traceless, and $c_0=c^{\alpha}_{\phantom{\alpha}\alpha}$, we can write the quadratic contribution as
\bea
\Gamma_{a,1}^{(1)}=-\frac{1}{32\pi^{2}}(\frac{1}{\epsilon}+\ln\frac{m^2}{\mu^2})\int d^{4}x\left(\frac{1}{6}c_0F_{\mu\nu}F^{\mu\nu}-\frac{1}{3}\tilde{c}_{\,\,\,\nu}^{\mu}F_{\mu\rho}F^{\rho\nu}\right),
\eea
so, if the $c^{\mu\nu}$ is traceless, our quadratic contributions (both divergent and finite one) are reduced to the aether-like form $\tilde{c}_{\,\,\,\nu}^{\mu}F_{\mu\rho}F^{\rho\nu}$.  We note that the aether-like term $\tilde{c}_{\,\,\,\nu}^{\mu}F_{\mu\rho}F^{\rho\nu}$ is accompanied by the factor $1/96$. Just this factor arises in our previous paper \cite{scalQED2a}, which confirms the validity of our result.

The quartic contribution to the EH effective action is finite as it must be and equal to
\begin{eqnarray}
\Gamma_{b,1}^{(1)}&=&\frac{1}{2304\pi^{2}m^4} \int d^{4}x\left[c_{\,\,\,\alpha}^{\alpha}\left(\frac{1}{5}F_{\mu\sigma}F^{\sigma\rho}F_{\rho\nu}F^{\nu\mu}+\frac{1}{4}\left(F_{\mu\nu}F^{\mu\nu}\right)^{2}\right) \right. \nonumber \\
&+& 2 \left. \left(F_{\mu\nu}F^{\mu\nu}\right)\left(c^{\sigma\rho}F_{\sigma\lambda}F_{\,\,\rho}^{\lambda}\right)
-  \frac{192}{95}\,c^{\mu\nu}F_{\mu\sigma}F^{\sigma\rho}F_{\rho\lambda}F_{\,\,\,\nu}^{\lambda}\right].
\end{eqnarray}
In principle, this result can be used for calculating the scattering amplitudes.

\section{EH effective action in the CPT-odd LV QED}

Now, let us proceed with the CPT-odd version of LV QED. In this case we have the theory
\bea
{\cal L}=(D_{\mu}\phi)^{\dagger}(\eta^{\mu\nu}+c^{\mu\nu})D_{\nu}\phi+u^{\mu}(\phi^*D_{\mu}\phi-(\phi D_{\mu}\phi)^*)-m^2\phi\phi^*-V(\phi\phi^*).
\eea
The corresponding one-loop effective action is
\bea
\Gamma^{(1)}[A]=i{\rm Tr}\ln (-D^2+2u^{\mu}D_{\mu}-m^2).
\eea
It is clear that the first nontrivial LV contribution depending only on the stress tensor itself but not on its derivatives will involve the second order in $u^{\mu}$. So, we expand our trace and arrive at
\bea
\Gamma^{(1)}_{LV,2}[A]=4i{\rm Tr}\frac{1}{D^2+m^2}u^{\mu} D_{\mu}\frac{1}{D^2+m^2}u^{\nu} D_{\nu}.
\eea
Now, we employ the Schwinger representations for two denominators:
\bea
\Gamma^{(1)}_{LV,2}[A]=4iu^{\mu}u^{\nu}{\rm Tr}\int_0^{\infty}ds dt e^{-(s+t)m^2} e^{-sD^2} D_{\mu}e^{-tD^2} D_{\nu}.
\eea
To proceed with this expression, we follow the same lines as in \cite{ourproperLV}, i.e. we insert the unit in the form 
$1=e^{sD^2}e^{-sD^2}$ and find
\bea
\Gamma^{(1)}_{LV,2}[A]=4iu^{\mu}u^{\nu}{\rm Tr}\int_0^{\infty}ds dt e^{-(s+t)m^2} D_{\nu}e^{-sD^2} D_{\mu}e^{sD^2}e^{-(s+t)D^2},
\eea
where we employed also the cyclic identity, and since (cf. \cite{McA,ourproperLV})
$$
e^{-sD^2} D_{\mu}e^{sD^2}=\exp(-2iesF)_{\mu\lambda}D^{\lambda},
$$
and assuming $[D_{\mu},F_{\nu\lambda}]=0$ (i.e. the stress tensor is constant), we have
\bea
\Gamma^{(1)}_{LV,2}[A]=4iu^{\mu}u^{\nu}{\rm Tr}\int_0^{\infty}ds dt e^{-(s+t)m^2} \exp(-2iesF)_{\mu\lambda} 
D_{\nu}D^{\lambda}e^{-(s+t)D^2}.
\eea
Then, we employ the formula (\ref{trace2d}) to obtain the trace for $D_{\nu}D^{\lambda}e^{-(s+t)D^2}$. As a result, we arrive at
\bea
\Gamma^{(1)}_{LV,2}[A]&=&4iu^{\mu}u^{\nu}\int_0^{\infty} \frac{ds dt}{16\pi ^2(s+t)^2} e^{-(s+t)m^2} \exp(-2iesF)_{\mu\lambda} 
\left(\frac{-ieF}{e^{-2ie(t+s)F}-1}\right)_{\phantom{\lambda}\nu}^{\lambda}\times\nonumber\\ &\times&
{\rm det}^{1/2}\left(\frac{ie(t+s)F}{\sinh (ie(t+s)F)}\right).
\eea
So, we  obtained the complete one-loop EH effective action in the form of some function of stress tensors. Again, terms up to fourth order can be read off from the power expansion of this formula as
\begin{equation}
-i\Gamma_{LV,2}^{(1)}[A]=e^{2}\Gamma_{a,2}^{(1)}+e^{4}\Gamma_{b,2}^{(1)},
\end{equation}
where
\begin{equation}
\Gamma_{a,2}^{(1)}=-\frac{1}{4\pi^{2}}\int d^{4}x\left[\frac{I_{3}(m^{2})}{24}u^{\alpha}u_{\alpha}F_{\mu\nu}F^{\mu\nu}+I_{4}(m^{2})u^{\mu}u_{\nu}F_{\mu\rho}F^{\rho\nu}\right],
\end{equation}
and
\begin{eqnarray}
\Gamma_{b,2}^{(1)}&=&\frac{u^{\alpha}u_{\alpha}}{576\pi^{2}}I_{5}(m^{2})\int d^{4}x\left[\frac{1}{5}F_{\mu\sigma}F^{\sigma\rho}F_{\rho\nu}F^{\nu\mu}+\frac{1}{4}\left(F_{\mu\nu}F^{\mu\nu}\right)^{2}\right]\nonumber\\
&+&\frac{u^{\sigma}u^{\rho}}{48\pi^{2}}I_{6}(m^{2})\int d^{4}x\left(F_{\mu\nu}F^{\mu\nu}\right)\left(F_{\sigma\lambda}F_{\,\,\rho}^{\lambda}\right)\nonumber\\
&+&\frac{u^{\mu}u^{\nu}}{12\pi^{2}}I_{7}(m^{2})\int d^{4}xF_{\mu\sigma}F^{\sigma\rho}F_{\rho\lambda}F_{\,\,\,\nu}^{\lambda},
\end{eqnarray}
with
\begin{eqnarray}
I_{3}(m^{2})&=&-\int_{0}^{\infty}dsdt\frac{e^{-(s+t)m^{2}}}{\left(s+t\right)},\\
%I_{3}(m^{2})&=&\int_{0}^{\infty}dsdt\frac{e^{-(s+t)m^{2}}}{\left(s+t\right)},\\
I_{4}(m^{2})&=&\int_{0}^{\infty}dsdt\frac{e^{-(s+t)m^{2}}}{\left(s+t\right)}\left[\frac{s^{2}}{\left(s+t\right)^{2}}-\frac{s}{(s+t)}+\frac{1}{6}\right]
%I_{4}(m^{2})&=& \int_{0}^{\infty}\frac{e^{-(s+t)m^{2}}}{(s+t)}\left[\frac{s^{2}}{\left(s+t\right)^{2}}+\frac{s}{(s+t)}+\frac{1}{6}\right]
,\\
I_{5}(m^{2})&=&\int_{0}^{\infty}dsdte^{-(s+t)m^{2}}(s+t),\\
I_{6}(m^{2})&=&\int_{0}^{\infty}dsdte^{-(s+t)m^{2}}(s+t)\left[\frac{s^{2}}{\left(s+t\right)^{2}}+\frac{s}{\left(s+t\right)}+\frac{1}{6}\right],\\
I_{7}(m^{2})&=&\int_{0}^{\infty}dsdte^{-(s+t)m^{2}}\left(s+t\right)\left[\frac{s^{4}}{\left(s+t\right)^{4}}-\frac{2s^{3}}{(s+t)^{3}}+\frac{s^{2}}{\left(s+t\right)^{2}}-\frac{1}{30}\right]
%I_{7}(m^{2})&=&\int_{0}^{\infty}dsdte^{-(s+t)m^{2}}(s+t)\left[\frac{s^{4}}{\left(s+t\right)^{4}}+\frac{2s^{3}}{(s+t)^{3}}+\frac{s^{2}}{(s+t)^{2}}-\frac{1}{30}\right].
\end{eqnarray}
All these integrals are finite as it must be by dimensional reasons. Doing the change of variables $s=ux,t=u(1-x)$, with $u\in [0,\infty]$, and $x\in[0,1]$,  we find $ds dt=  u du dx$,  and $s+t=u$, we easily calculate the integrals, actually,  we have $I_3= - \frac{1}{m^2}$,    $I_5 = I_6  = \frac{2}{m^{6}} $ and $I_4 = I_7 = 0$. 
So, explicitly we can write the quadratic contribution in the form
\begin{equation}
\Gamma_{a,2}^{(1)}= \frac{u^{\alpha}u_{\alpha}}{96\pi^{2}m^2}\int d^{4}x F_{\mu\nu}F^{\mu\nu},
\end{equation}
so,  we have the Lorentz invariant, Maxwell-like  contribution. 

The quartic contribution to the EH effective action in this case looks like
\begin{eqnarray}
\Gamma_{b,2}^{(1)}&=&\frac{u^{\alpha}u_{\alpha}}{288\pi^{2}m^6}\int d^{4}x\left[\frac{1}{5}F_{\mu\sigma}F^{\sigma\rho}F_{\rho\nu}F^{\nu\mu}+\frac{1}{4}\left(F_{\mu\nu}F^{\mu\nu}\right)^{2}\right]\nonumber\\
&+&\frac{u^{\sigma}u^{\rho}}{24\pi^{2}m^6}\int d^{4}x\left(F_{\mu\nu}F^{\mu\nu}\right)\left(F_{\sigma\lambda}F_{\,\,\rho}^{\lambda}\right).
\end{eqnarray}
We find that we obtained two known quartic invariants arising within Lorentz-invariant calculations of the EH effective action, and only one of only possible quartic LV terms. In principle, these results can be used for studying of scattering processes, similarly to \cite{Fur}.

\section{Summary}

We calculated the lower LV contributions to the EH effective action in scalar QED, both in CPT-even and CPT-odd cases. We found that the lower result in the CPT-even case is of the first order in the LV parameter, while in the CPT-odd case -- of the second order. 
In principle, it could mean that the LV contributions to the EH effective action in the CPT-odd theory are suppressed in comparison with those ones in the CPT-even case, unless the mass of the fermion is not very {\bf tiny}.  We can argue this as follows. Within our calculations, we follow the perturbative approach, within which, the LV parameters are assumed to be very small (f.e., following \cite{DataTables}, the typical estimation for the  $c^{\mu\nu}$ is about $10^{-15}$, and for  the vector coefficient $u^{\mu}$, within various estimations, is no more than $10^{-15}$ GeV). Actually, to argue that the contribution from the CPT-even sector is more relevant, besides the evident fact that the second-order contribution in the perturbative series is less important than the first one, we can use the fact that, for example, the aether-like contribution $a^{\mu}a_{\nu}F_{\mu\lambda}F^{\nu\lambda}$, being generated by the CPT-even LV term, after subtracting the divergences, looks like $\frac{1}{96\pi^2}\ln\frac{m^2}{\mu^2}\tilde{c}^{\mu}_{\nu}F_{\mu\rho}F^{\rho\nu}$, and being generated by the CPT-odd LV term, looks llke $\frac{1}{4\pi^2m^2}u^{\mu}u_{\nu}F_{\mu\lambda}F^{\nu\lambda}$ (for the traceless $\tilde{c}^{\mu\nu}$ and light-like $u^{\mu}$ respectively), and, for our estimations for $c^{\mu\nu}$ and $u^{\mu}$, and  it is clear that $|\frac{u^{\mu}u^{\nu}}{m^2}|\ll |\tilde{c}^{\mu\nu}|$, thus, the contribution proportional to $u^{\mu}u^{\nu}$ is really suppressed.  

Certainly, our results can be applied for studies of loop corrections to photon-photon scattering in LV QED. It is clear that the generalization of these results to the non-Abelian case is straightforward since the dependence of the effective action on field strengths is completely described by the commutators of covariant derivatives.

A possible continuation of our study could consist in studying the EH effective action in other possible LV extensions of the scalar QED, including non-minimal ones. We plan to perform this study in one of forthcoming papers.

\acknowledgments

The work of A. Yu.\ P. has been partially supported by the CNPq project No. 303777/2023-0. The work of A. C. L. has been partially supported by the CNPq project No. 404310/2023-0.

%%%%%%%%%

\end{document}